\documentclass[aps,pre,reprint,superscriptaddress,amsmath,amssymb]{revtex4-2}

\usepackage{graphicx}
\usepackage{amsmath}
\usepackage[colorlinks=true,allcolors=blue]{hyperref}

\newcommand{\nn}{\nonumber}

\newcommand{\veps}{\varepsilon}
\newcommand{\ea}{eqnarray}
\newcommand{\tl}{\tilde}

\begin{document}

\title{Theory of magneto-electrokinetics in unidirectional electrolyte flow:
Hydrodynamic potential and Hall effect}

\author{Yoav Tsori}
\email{tsori@bgu.ac.il}
\affiliation{Department of Chemical Engineering, Ben-Gurion University of the Negev,
Beer-Sheva 84105, Israel}

\date{\today}

\begin{abstract}

We develop a theoretical framework for magneto-electrokinetic transport in dilute 
electrolytes under unidirectional flow and streaming current conditions. Unlike 
standard magnetohydrodynamics in
liquid metals, where charge carriers are electrons and conductivity is typically 
uniform, electrolytes exhibit spatially varying ion densities and additional forces such as 
osmotic pressure. In steady state, the coupled Navier--Stokes, Nernst--Planck, and 
Poisson equations become analytically tractable. We derive a Boltzmann-like expression for 
ion distributions that incorporates both the electrostatic potential and a novel 
hydrodynamic potential arising from the magnetic field. This potential depends explicitly 
on the flow profile and is not screened close to the bounding walls. Applying the theory to 
Poiseuille flow, we derive expressions for the induced Hall potential and demonstrate its 
near-linear dependence on flow velocity and magnetic field strength, while remaining largely 
independent of salt concentration. To validate this one-dimensional framework, we analyze the 
two-dimensional spatial transients near the channel entrance. We demonstrate a 
separation of length scales: the Hall electric field and space charge adjust to the magnetic field 
almost instantaneously, whereas the macroscopic salt concentration wake develops over a much 
longer Graetz length dictated by the P\'eclet number. We also examine a similar flow with 
counterions only and no added salt, determining the resulting potential and ion distributions. This 
work offers new theoretical tools for understanding ion transport in microfluidic and 
electrochemical systems.

\end{abstract}

\maketitle


\section{Introduction}

Magnetic fields play a pivotal role in a wide range of scientific and engineering 
domains. From fusion reactors and metallurgical processes to microfluidic devices and 
electrochemical systems to plasma accelerators and astrophysics, the Lorentz force facilitates 
controlling fluid motion, enhancing mass and heat transfer, mixing and vortex manipulation, and 
mitigating instabilities \citep{bau_magnetochemistry_2022,vilagines_rjaset_2012,
bau_mech_res_comm_2009}. The Lorentz force can induce complex flow patterns, modify 
boundary layer behavior, and influence 
electrochemical reaction kinetics \citep{bau_magnetochemistry_2022,kuhn_coe_2023}.

In electrochemical systems, magnetic fields have been harnessed to improve electrodeposition 
uniformity, suppress dendrite formation in batteries, and enhance mixing in microreactors.
\citep{li_jmca_2022,kuhn_coe_2023} For instance, the application of a static magnetic field in 
zinc-air batteries has been shown to regulate ion transport and reduce surface roughness, thereby 
extending cycle life \citep{li_jmca_2022}. Similarly, magnetic-driven flows in microfluidic 
channels enable non-mechanical fluid control, which is critical for lab-on-a-chip technologies and 
biomedical diagnostics 
\citep{bau_magnetochemistry_2022,dong_nfpm_2019,tse_adv_func_mat_2016,ong_bpee_2023}.

Realistic flows are two- or three-dimensional and complex in nature and usually lack analytical 
solutions and even numerical ones are difficult to obtain. 
Here, we restrict ourselves to one-dimensional flows in magnetic fields, which offer
better physical insight and are easier to analyze analytically. We consider steady unidirectional 
flow of dilute electrolytes in a perpendicular constant magnetic field. 
In standard magnetohydrodynamics for liquid metals, conductivity is typically uniform, and the 
highly mobile electrons lead to quasi-neutrality, effectively shrinking space-charge boundary 
layers to atomic scales \citep{muller_book,landau_edcm, melcher_1981, probstein_2003}. 
Here, dissolved ions drift with the 
embedding liquid and, in addition, experience a Lorentz force acting in the perpendicular 
direction. The ions' density thus varies in space, leading to spatially non-uniform conductivity. In 
this system, the Lorentz and electric forces and the ions' osmotic pressure affect the fluid flow, 
while the ionic density is dictated by their diffusion, drift with the flow, and the Lorentz force. 

The assumption of unidirectional flow allows us to obtain analytical solutions to both the 
Navier-Stokes equation and the ionic density. Surprisingly, the Navier-Stokes equation sequentially 
decouples from the ions. The ionic density obeys a Boltzmann-like 
distribution depending on a hydrodynamic potential $U_{\rm hyd}$ and the electrostatic potential. 
We apply this general result to a ``Hall effect'' in flowing electrolytes in channel flow to find the Hall 
potential and liquid velocity profile.

The remainder of the paper is organized as follows. In Section \ref{sec_theory}, we formulate the 
general equations governing magneto-electrokinetic transport for a unidirectional flow. In Section 
\ref{sec_poiseuille}, we apply this framework to a salt-containing electrolyte in Hagen-Poiseuille 
flow, deriving the governing Poisson-Boltzmann equation and evaluating the induced transverse 
Hall potential. To rigorously justify the fully developed one-dimensional assumption, Section 
\ref{sec:entrance_effect} details the 
two-dimensional spatial transients near the channel entrance. Section \ref{hall_ci} extends the 
analysis to a counterion-only electrolyte driven between charged plates. Finally, concluding 
remarks are provided in Section \ref{sec:conclusions}.

\section{Magneto-electrokinetics of dilute electrolytes in unidirectional flow}\label{sec_theory}

\begin{figure}
\centering
\includegraphics[width=0.5\columnwidth,bb=140 90 400 500,clip]{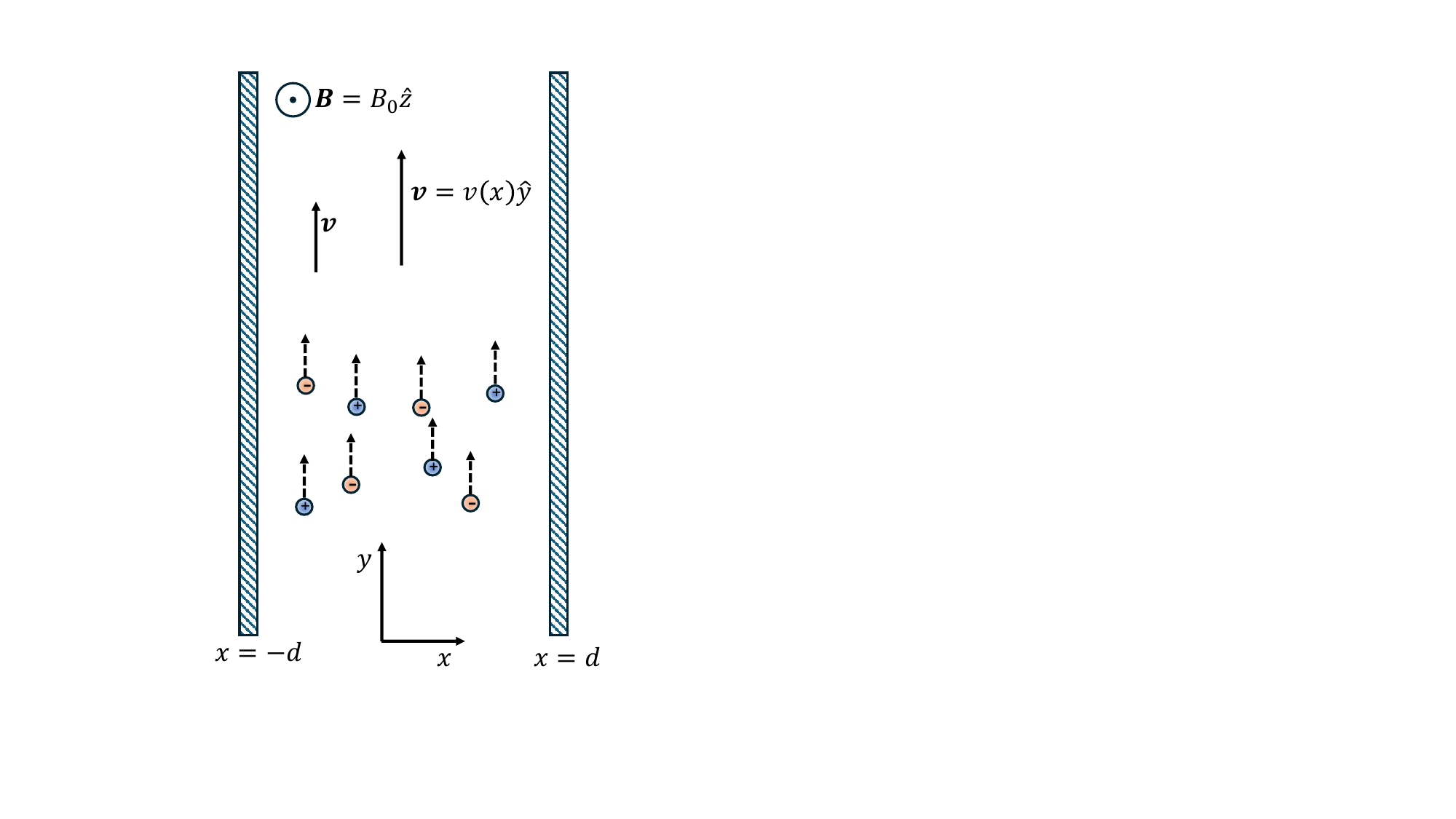}
\caption{Illustration of electrolytes flowing in a slit under a perpendicular magnetic field. The 
electrolytes is confined by two flat solid surfaces at $x=\pm d$. Flow is unidirectional along $y$, 
while the external magnetic field is ${\bf B}=B_0\hat{z}$. Cations and anions (colored spheres) are 
dissolved in the carrying liquid. Their number densities are $n^+$ and $n^-$, respectively.
}
\label{fig1_illus}
\end{figure}

We consider a dilute electrolyte flowing past one or more flat and smooth surfaces and subject to 
a perpendicular magnetic field. The coordinate system is such that $\hat{x}$ is perpendicular to 
the surface, the flow is parallel to $\hat{y}$, and the magnetic field is ${\bf B}=B_0\hat{z}$ (see Fig. 
\ref{fig1_illus}). The ions dissolved in the liquid are assumed to be point-like and symmetric; their 
charge and number densities are $q$ and $n^+$ for the cations and $-q$ and $n^-$ for the anions. 
The incompressibility and Navier-Stokes equations for the pressure $p$ and flow velocity ${\bf v}$ are
\begin{\ea}
\nabla\cdot{\bf v}&=&0,\\
\rho\left[\frac{\partial {\bf v}}{\partial t}+({\bf v}\cdot{\bf \nabla}){\bf
v}\right]&=&-\nabla p+\eta\nabla^2{\bf v}+\rho_e{\bf E}\nn\\
&+&{\bf j}\times{\bf B}-k_BT\nabla (n^++n^-).
\label{navier_stokes}
\end{\ea}
Here, $\rho$ is the liquid's mass density, $\eta$ its viscosity, $\rho_e=q(n^+-n^-)$ the charge 
density, ${\bf j}$ the electric current density, ${\bf E}$ the electric field, $k_B$ the Boltzmann 
constant and $T$ the absolute temperature. The sum $\rho_e{\bf E}+{\bf j}\times{\bf B}$ on the 
right-hand side of Eq. (\ref{navier_stokes}) is due to the Lorentz force on moving charged 
ions,\citep{sader_prb_2025} while the last term is an osmotic pressure of dilute ideal gas of two 
ions with densities $n^+$ and $n^-$. The electric field is not imposed by external potentials; rather, 
it arises self-consistently from nonuniform ion separation. We
employ the so-called streaming current conditions, where the component of the field along
the flow vanishes. Experimentally, this corresponds to a steady, continuous bulk fluid pumping by 
a pressure gradient through a channel with non-conducting (insulating) walls, which prevents 
external short-circuiting of the induced Hall current.

In the Nernst-Planck framework, the ionic densities obey the continuity equation
\begin{\ea}\label{np}
\frac{\partial n^\pm}{\partial t}+\nabla\cdot {\bf h}^\pm=0.
\end{\ea} 
The ionic fluxes ${\bf h}^\pm={\bf h}_{\rm L}^\pm+{\bf h}_{\rm d}^\pm+{\bf h}_{\rm adv}^\pm$ 
have three contributions: 
\begin{\ea}
{\bf h}^\pm_{\rm L}&=&\pm n^\pm \mu q\left({\bf 
E}+{\bf v}\times {\bf B}\right),\nn\\
{\bf h}^\pm_{\rm d}&=&-D\nabla n^\pm,\nn\\
{\bf h}^\pm_{\rm adv}&=&n^\pm{\bf v}.\label{fluxes}
\end{\ea} 
The first one, ${\bf h}_{\rm L}$, is due to the Lorentz force ${\bf f_{\rm 
L}}=\pm q\left({\bf E}+{\bf v}\times {\bf B}\right)$ acting on the charges. The mobility $\mu$, assumed 
equal for both ion types, obeys the Einstein relation $\mu=D/k_BT$, where $D$ is the diffusion 
constant \citep{friedman_jpc_1965,friedman_jcp_1987,felderhof_1988,felderhof_1990}.  
The second term, ${\bf h}_{\rm d}$, is a diffusive flux of an ideal gas of ions. ${\bf 
h}^\pm_{\rm adv}$ is a flux for advection with the flow.

For a system that is translationally invariant along $y$ (i.e., neglecting edge effects)
and in unidirectional flow, the velocity and ion density depend on $x$ only:  
${\bf v}=v(x)\hat{y}$ and $n^\pm=n^\pm(x)$. Ions are carried by the liquid in the $y$ 
direction and feel a perpendicular magnetic-origin force $q{\bf v}\times{\bf B}$ in the $x$ direction. 
Their number density is nonuniform in $x$, leading to an electric field ${\bf E}=E(x)\hat{x}$
and osmotic pressure opposing 
the magnetic force. In steady state and in the absence of an electrochemical reaction at the bounding 
surfaces,  these forces exactly cancel each other. Formally, the Nernst-Planck equation 
becomes $\nabla\cdot{\bf h}^\pm=0$. Since ${\bf h}^\pm$ are independent of $y$, we find
\begin{\ea}\label{zero_flux}
\pm n^\pm \mu q\left({\bf E}+{\bf v}\times {\bf B}\right)-D\nabla n^\pm=0.
\end{\ea} 

In the Navier-Stokes equation, the term $({\bf v}\cdot{\bf \nabla}){\bf v}=0$ {\it vanishes identically} 
as 
long as we are before turbulence, i.e., when the Reynolds number $R_e$ is sufficiently small. A 
tremendous simplification arises because the Lorentz and 
diffusive fluxes are oriented along $x$ -- all body forces in the $y$ component vanish and
Eq. (\ref{navier_stokes}) reduces to the Stokes equation:
\begin{\ea}\label{eq_stokes}
-\frac{\partial p}{\partial y}+\eta\frac{\partial^2 v(x)}{\partial x^2}=0
\end{\ea} 
as if there were no magnetic or electric forces. There is a unique flow profile $v(x)$ depending on the 
pressure gradient and other boundary conditions of the problem. 
In Sec. \ref{sec_poiseuille} we employ the specific Hagen-Poiseuille flow but at this point we do not 
use it. 

We have achieved sequential decoupling between the flow profile and the ionic densities -- once $v(x)$ 
is known from Eq. (\ref{eq_stokes}), it can be substituted in Eq. (\ref{zero_flux}) to obtain:
\begin{\ea}\label{eq_ions_boltzmann}
n^\pm(x)&=&n_m\exp\left(\frac{\mp q\psi\pm U_{\rm hyd}(x)}{k_BT}\right),
\end{\ea} 
where
\begin{\ea}\label{eq_Uhyd}
U_{\rm hyd}(x)&\equiv &q\int_0^x v(x')B_0dx'
\end{\ea} 
is a magnetic-field-dependent ``hydrodynamic potential'' that ions feel and $n_m$ is a constant related 
to conservation of the total ion number. 

The hydrodynamic potential $U_{\rm hyd}(x)$ introduced in Eq. (\ref{eq_Uhyd}) contains the 
cumulative effect of the Lorentz force acting on ions as they are advected by the flow. Physically, it 
represents the work done by the magnetic field on a charged particle moving along the flow 
direction. Conceptually analogous non-equilibrium effective drift potentials have also been derived for 
externally driven soft-matter systems \citep{zaccone_pre_2013}. The potential $U_{\rm hyd}$ 
depends explicitly on the velocity profile and is nonzero only in the presence of both flow and 
magnetic field. It introduces a directional bias in ion distribution; in the bulk, the magnetic and 
electric components of the Lorentz force cancel, allowing the electrostatic field to persist far from 
the boundary walls. This mechanism is central to the emergence of the Hall potential in flowing 
electrolytes and marks a departure from classical equilibrium-based screening models.

The electrostatic potential $\psi$ satisfies the Poisson's equation  
$\nabla^2\psi=-(q/\veps)\left(n^+-n^-\right)$. Substitution of $n^\pm$ from Eq. 
(\ref{eq_ions_boltzmann}) then yields the following Poisson-Boltzmann equation
\begin{\ea}\label{eq_pb}
\psi''&=&\frac{2qn_m}{\veps}\sinh\left(\frac{q\psi-U_{\rm 
hyd}(x)}{kT}\right).
\end{\ea} 

Note that even though the ions are not in thermodynamic equilibrium, they obey a Boltzmann-like 
distribution with a non-electrostatic contribution, $U_{\rm hyd}$, depending on the flow 
profile. The term $U_{\rm hyd}$ appears due to the presence of magnetic field and vanishes in its 
absence. A second important point is that ions are {\it not} classically screened. They feel a 
long-range force whose range is dictated by the flow profile and is usually algebraic in the distance 
from the wall. This means the ``electric double layer'' in this problem is different from the usual 
one; anions and cations can be separated over large distances. The magnetic field does not 
typically exert strong forces (see estimate below), so actual differences in anion and cation density 
may not be large. Nonetheless, the length-scale characterizing the variations in $n^\pm(x)$ can be 
significantly larger than the Debye screening length. 

The boundary conditions for $\psi$ depend on the physical realization. In this work, 
we assume no imposed external potential across the channel. The potential results from the influence 
of the magnetic field solely, or due to surface charges on channels' walls. This means that $\psi'$ on 
the walls is zero or constant, respectively. In Secs. \ref{sec_poiseuille} and \ref{hall_ci} we employ 
specific 
boundary conditions appropriate to the physical realization.

Before we look in detail on a specific flow problem, let us verify the validity of the assumption that 
the total magnetic field ${\bf B}$ equals the external imposed field $B_0\hat{z}$. The assumption 
should hold if the field produced by the moving ions is sufficiently small. We start with the 
Maxwell's equations:
\begin{\ea}
\nabla\times{\bf E}&=&-\frac{\partial{\bf B}}{\partial t},~~~\hspace{2cm}{\bf 
B}=\mu_0{\bf H},\nn\\
\nabla\times{\bf H}&=&{\bf j}+\veps\frac{\partial{\bf E}}{\partial 
t},\hspace{2cm}\nabla\cdot{\bf B}=0.
\end{\ea}
Here, $\mu_0$ is the vacuum permeability and $\veps$ is the permittivity of the liquid. In our 
problem, 
the electrical conductivity is not constant and the self-consistent expression for the electrical 
current density is ${\bf j}=q(n^+-n^-)v(x)\hat{y}$. 

From Amp\`{e}re's law with Maxwell's correction,
\begin{\ea}
\frac{1}{\mu_0}\nabla\times\left(\nabla\times{\bf 
B}\right)&=&\nabla\times\left(q(n^+-n^-){\bf 
v}\right).\nn
\end{\ea}
The operator identity $\nabla\times\left(\nabla\times{\bf B}\right)=\nabla\left(\nabla\cdot{\bf 
B}\right)-\nabla^2{\bf B}$ allows to write 
\begin{\ea}\label{eq_B}
-\frac{1}{\mu_0}\nabla^2{\bf B}=\nabla\times\left(q(n^+-n^-){\bf v}\right).
\end{\ea}
Parameter $n$ depends on $x$ only and $\nabla\times {\bf v}=v'(x)\hat{z}$, and therefore Eq. 
(\ref{eq_B}) has only a $z$ component. We write the magnetic field as ${\bf B}=B_0\hat{z}+{\bf b}$ 
where ${\bf b}=b\hat{z}$ is the correction to the external field due to the moving ions. Equation 
(\ref{eq_B}) can be written as $\nabla^2 b=-\mu_0q\partial_x((n^+-n^-)v)$. 

One can estimate $b$ from $\partial b/\partial x\sim \mu_0qnv(x)$. For a typical flow velocity 
$v=0.1$ m/s and average ion  density of $n_0=1$ mM, with monovalent ions of $q=e$ ($e$ is the 
elementary charge), we find
\begin{\ea}
\frac{\partial b}{\partial x}\sim 10^{-2} {\rm ~T/m}
\end{\ea}
or smaller, because in most locations $|n^+-n^-|\ll n_0$. For small enough systems and large 
external magnetic fields ($B_0\sim 1$ T), $b\ll B_0$ and we may thus assume that ${\bf 
B}=B_0\hat{z}$ is constant in space. With increasing values of the external field $B_0$, $b$ 
becomes even smaller in comparison with $B_0$. In concentrated electrolytes or ionic liquids 
where $n_0\sim 1$M and for channel width $\sim 10$ cm, $b$ becomes comparable to $B_0$, and 
cannot be neglected. Higher flow rates, with velocities $v_0$ of the order of several meters per 
second would also increase $b$, but then the assumption of laminar flow does not hold.

In the next section we apply the formalism expressed in Eqs. (\ref{eq_stokes}), 
(\ref{eq_ions_boltzmann}), (\ref{eq_Uhyd}) and (\ref{eq_pb}) to the ``Hall effect'' in flowing 
electrolytes.

\section{Hall effect in flowing electrolytes in Poiseuille flow}\label{sec_poiseuille}

The discussion and derivation above apply to general unidirectional flows. One such flow occurs in 
the Hall effect in flowing electrolytes. Soon after the discovery of the Hall effect in 1879, there were 
attempts to realize it in electrolytes. Most early experiments failed or were heavily criticized 
for flaws. 
\citep{bagard_cr_1896,bagard_nc_1898,moretto_nc_1902,delvalez_jpta_1909,oxley_prsla_1913} 
An early theory by Donnan assumed anions and cations are advected in opposite directions due to 
a parallel electric field, hence yielding a weak Hall effect \citep{donnan_proc_phys_soc_1897}. Years 
later, theoretical works dealt primarily with the calculation of the ions' mobility in given crossed 
electric and magnetic fields 
\citep{friedman_jpc_1965,friedman_jcp_1987,felderhof_1988,felderhof_1990}. 
It is important to distinguish fluid motion driven by a parallel electric 
field from our hydrodynamic scenario: when cations and anions migrate in opposite directions, 
the transverse Lorentz force pushes both ion types in the \textit{same} direction, 
leading to local salt accumulation but no leading-order charge separation. In 
contrast, in the pressure-driven hydrodynamic flow studied here, all ions are 
advected in the same direction. The Lorentz force therefore pushes cations and 
anions in \textit{opposite} transverse directions, actively inducing a 
charge separation and the resulting Hall potential.

Starting in 1975 and led by the Japanese school \citep{aogaki_1975a,aogaki_1975b}, researchers 
investigated the Hall effect in the context of electrochemical processes. These works examine ionic 
currents flowing {\it perpendicular} to the confining surfaces. The Lorentz force is thus {\it parallel} 
to the flow. Most of these studies do not solve the Poisson's equation. Besides electrochemistry, 
\citep{fahidy_jae_1983,bau_phys_fluids_2005} modern works focus on magnetohydrodynamic 
pumping, where the Lorentz force acts parallel to a channel's 
walls,\citep{de_rooij_2005,chakra_jpdap_2006,bau_micro_nanofluids_2011,bau_phys_fluids_2012}
or on heat transfer \citep{hassan_cjp_2002,hassan_amm_2003}.

We now look in detail at Hagen-Poiseuille flow of an electrolyte in a slit between two surfaces and 
subject to a perpendicular magnetic field. Two parallel and smooth surfaces are at $x=\pm d$. 
Their length in the $y$ direction, $L$, is assumed sufficiently long to neglect edge effects. A 
pressure difference $\Delta p$ is applied along the slit, and the solution to Eq. 
(\ref{eq_stokes}) for laminar flow is the familiar parabolic profile
\begin{\ea}\label{parabolic_flow}
{\bf v}(x)&=&v_0\left(1-\left(\frac{x}{d}\right)^2\right)\hat{y},\\
v_0&=&\frac{\Delta p d^2}{2\eta L}.\nn
\end{\ea} 

The hydrodynamic potential integral in Eq. (\ref{eq_Uhyd}) for Hagen-Poiseuille flow
is 
\begin{\ea}
U_{\rm hyd}(x)=qv_0B_0\left(x-\frac13\frac{x^3}{d^2}\right).
\end{\ea} 
The confining surfaces are assumed electrically neutral. Possible materials that remain neutral 
in water are Teflon, polypropylene, polystyrene, and polyethylene. The electrostatic potential depends 
on $x$ only. One needs to solve Eq. (\ref{eq_pb}) subject to the condition $\psi'(x=\pm 
d)=0$, which follows the Gauss' theorem for uncharged surfaces. Experimentally, it corresponds to 
a channel made of, for example, plastic material, or a metal coated with thin layer of inert material.
Other experimental configurations could also be considered, such as imposing a constant 
potential: $\psi(x=\pm d)=\pm V/2$, where $V$ denotes the potential drop across the channel 
walls, thereby establishing an average external electric field. Under the no-charge conditions 
assumed here, the electric field arises solely from the magnetic field and the motion of charges.

The potential $\psi$ is an anti-symmetric function of $x$. The constant $n_m$ appearing in Eq. 
(\ref{eq_pb}) can now be interpreted as the yet-unknown midplane concentration, $n_m=n(x=0)$. It 
is related to the average ion density $n_0$ via the material conservation integral
\begin{\ea}
\int_{-d}^d n^\pm(x)dx=2n_0 d.
\end{\ea} 

We now define the dimensionless field $\tl{B}$, electrostatic potential $\tl{\psi}$, and Debye length 
$\lambda_D$ by
\begin{\ea}\label{eq_dimensionless}
\tl{B}&=&\frac{qv_0B_0d}{k_BT}~~~~,~~~~~~\tl{\psi}=\frac{q\psi}{k_BT}~~~~,~~~~~
\lambda_D^2=\frac{\veps k_BT}{2n_0q^2}.
\end{\ea} 
Further, length is scaled by the Debye length:
\begin{equation}
\tl{x} = \frac{x}{\lambda_D} \quad , \quad \tl{d} = \frac{d}{\lambda_D}.
\end{equation}
This rescaling of length in the governing equations allows us to resolve 
the steep concentration gradients within the nanometer-scale space-charge 
region, explicitly isolating the boundary layer behavior from the macroscopic 
bulk flow. The definition of $\phi$ as
\begin{\ea}\label{eq_phi_psi}
\phi=\tl{\psi}-\tl{B}\left(\frac{\tl{x}}{\tl{d}}-\frac13\left(\frac{\tl{x}}{\tl{d}}\right)^3\right).
\end{\ea}
allows us to write the Poisson-Boltzmann equation and its boundary conditions as
\begin{\ea}\label{eq_poiseuille_salt}
\phi''&=&\lambda\sinh(\phi)+\frac{2\tl{B}\tl{x}}{\tl{d}^3},\\
\lambda\int_0^{\tl{d}}\cosh(\phi)d\tl{x}&=&\tl{d}~,~~~~~~~\phi(\tl{x}=0)=0~,~~~~~\phi'(\tl{x}
=\tl{d})=0,\nn
\end{\ea} 
where the prime denotes differentiation with respect to the dimensionless 
coordinate $\tl{x}$. Equation (\ref{eq_poiseuille_salt}) is an inhomogeneous 
ordinary differential equation with a location-dependent ``force'' term proportional to the 
magnetic field via $\tl{B}$.
The nonlinear eigenvalue $\lambda\equiv n_m/n_0$ has a unique value for which $\phi$ satisfies 
both boundary conditions {\it and} the conservation integral. The eigenvalue $\lambda$ quantifies 
the deviation of the midplane ion concentration $n_m$ from the average bulk concentration 
$n_0$. It is smaller than unity, indicating ion depletion in the midplane and accumulation near the 
walls. Since the local conductivity depends on the ionic concentration, $\lambda$ also serves as a 
proxy for the spatial variation in conductivity across the slit. In this sense, $\lambda$ encapsulates 
both electrostatic and hydrodynamic contributions to ion transport and can be linked to 
measurable quantities such as the transverse Hall potential and local current density.

Before we solve this equation, let us look at $\tl{B}$. This parameter is the ratio between the 
energy of the Lorentz force $\sim qvB$ integrated over the slit's width $d$ and the thermal energy 
$k_BT$. What is its magnitude? When $v_0=1$ cm/s, $q=2e$, $B_0=0.1$ T, $d=0.1$ m and at room 
temperature, $\tl{B}\simeq 0.008$. In microfluidic systems $d$ is smaller and $\tl{B}$ is 
correspondingly smaller. One can express $\tl{B}$ using the Reynolds number $Re\equiv \rho 
v_0d/\eta$ as $\tl{B}=MRe$, where the dimensionless magnetic number is $M\equiv q\eta B/\rho 
k_BT$. Thus, to increase $\tl{B}$ while keeping $Re$ fixed, one needs to increase $M$, that 
is, increase charge, viscosity or field, or reduce mass density or temperature. 

The term $2\tl{B}\tl{x}/\tl{d}^3$ in Eq. (\ref{eq_poiseuille_salt}) is absent from classical 
Poisson-Boltzmann formulations. Ions feel a long-range force whose range is dictated by the flow 
profile, usually algebraic in the distance from the wall. This means that ions are {\it not} 
exponentially screened and the ``electric double layer'' is very different from the usual one; 
anions and cations can be separated over large distance. We emphasize that the fundamental 
Debye screening length $\lambda_D$ remains an unchanged thermodynamic property of the 
electrolyte. While the magnetic field induces a macroscopic algebraic dependence in the bulk of 
the channel, the boundary layers at the walls remain exponential and are governed by the 
standard $\lambda_D$. The external force is not strong so actual differences in anion/cation 
density would not be large. Still, the length-scale is of the order of the 
conduit's size. One should be careful when looking at exceedingly small or large slit widths in Eq. 
(\ref{eq_poiseuille_salt}), because $\tl{B}=qv_0B_0d/kT$, and in parabolic flow $v_0\sim d^2$, Hence 
$\tl{B}/\tl{d}^3$ is independent of $d$.

When $\tl{B}/\tl{d}^2\ll 1$, Eq. (\ref{eq_poiseuille_salt}) can be linearized as 
$\phi''\approx\lambda\phi+2\tl{B}\tl{x}/\tl{d}^3$ yielding
\begin{\ea}\label{eq_phi_ana}
\phi=\frac{2\tl{B}}{\tl{d}^3\lambda}\frac{\sinh(\sqrt{\lambda}\tl{x})}{\sqrt{\lambda}
\cosh(\sqrt{\lambda}\tl{d})}-\frac{2\tl{B}}{\tl{d}^3\lambda}\tl{x}.
\end{\ea}
What is the Hall potential across the two sides of the slit? We anticipate $\lambda$ is slightly smaller 
than unity; when $\tl{d}\gg 1$ it follows that $\phi(\tl{d})\approx -2\tl{B}/\tl{d}^2$. Converted back to 
$\tl{\psi}$ values, the electrostatic potential is 
\begin{\ea}
\tl{\psi}(\tl{d})\approx \frac23\tl{B}-\frac{2\tl{B}}{\tl{d}^2\lambda}.
\end{\ea}
The dependence on salt content $n_0$ appears solely through the small second term,
$\tl{B}/\tl{d}^2$. When this term can be neglected compared with unity, $\tl{\psi}(\tl{d})\approx 
2\tl{B}/3$ is independent of salt content and one finds the potential drop $\psi_{\rm Hall}\equiv 
\psi(d)-\psi(-d)$ to be
\begin{\ea}\label{eq_psid_approx}
\psi_{\rm Hall}\approx \frac43v_0B_0d.
\end{\ea}
When $v_0=1$ cm/s, $B=1$ T, and $d=0.1$ m, one finds $\psi_{\rm Hall}\approx 1.3$ mV 
independent of the amount of salt. While this induced potential is seemingly small, it generates 
macroscopic, measurable effects. In fact, commercial electromagnetic flowmeters routinely and 
accurately infer fluid flux by measuring these exact transverse Hall potentials in the $1$--$10$ mV 
range in aqueous flows \citep{baker_flow_2016, emerson_magmeter, yokogawa_magmeter}. Our 
theory provides the foundational micro-physics for this well-established industrial phenomenon.

\begin{figure}
\centering
\includegraphics[width=\columnwidth,clip]{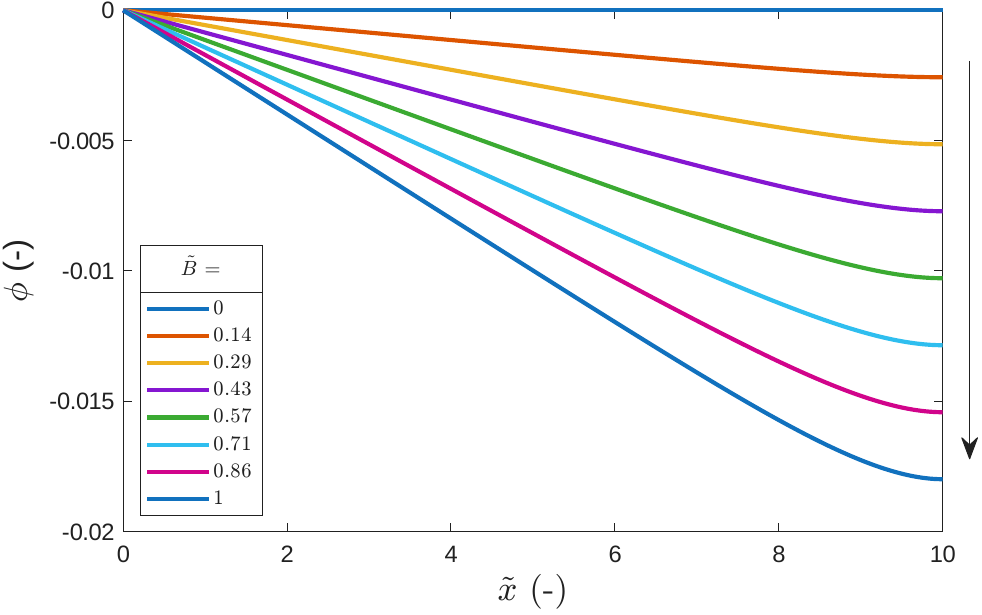}
\caption{Dimensionless potential profiles $\phi(\tl{x})$ for salt-containing electrolytes flowing in a 
slit under perpendicular magnetic field. They are numerical solutions of Eq. 
(\ref{eq_poiseuille_salt}) for $\tl{d}=10$. Each curve corresponds to a different value of the scaled 
magnetic field $\tl{B}$ (see legend), with the downward arrow indicating the direction of 
increasing $\tl{B}$. The profiles are antisymmetric about the midplane $\tilde{x} = 
0$, where the potential is zero. As $\tl{B}$ increases, the potential $\phi(\tl{d})$ becomes more 
negative, indicating stronger ion depletion in the midplane and accumulation near the 
boundaries. This behavior reflects long-range ion separation, in contrast to classical electric double 
layers with exponential decay.
}
\label{fig_phi_vs_x}
\end{figure}
\begin{figure}
\centering
\includegraphics[width=\columnwidth,clip]{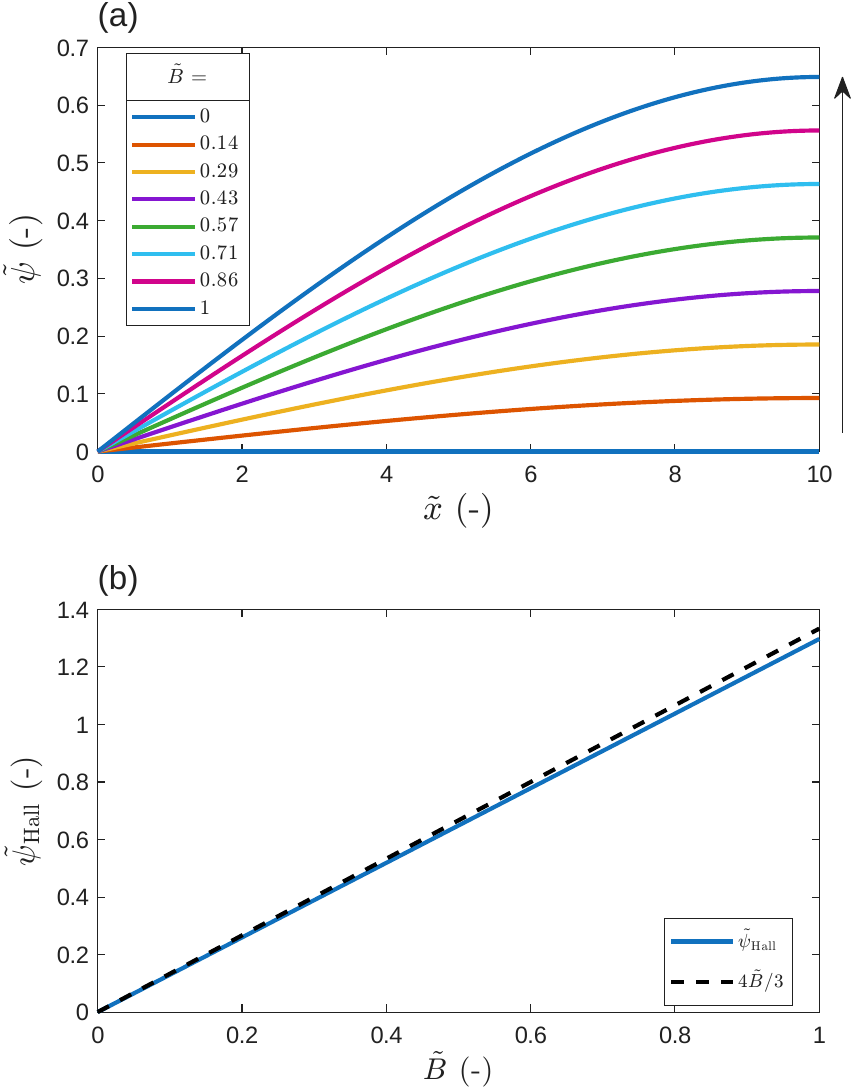}
\caption{(a) Dimensionless electrostatic potential $\tilde{\psi}(\tilde{x})$ for different values of the 
scaled magnetic field $\tl{B}$, derived from the potential $\phi(\tl{x})$ in Fig.~\ref{fig_phi_vs_x} 
using Eq. (\ref{eq_phi_psi}). The curves show how the potential increases away from the midplane, 
indicating a persistent electric field throughout the slit. The scaled magnetic field value is given by 
the legend (vertical arrow denotes increasing $\tl{B}$). (b) Dimensionless Hall potential 
$\tl{\psi}_{\rm Hall}=\tl{\psi}(\tl{d})-\tl{\psi}(\tl{-d})$ as a function of $\tl{B}$. The solid line shows 
numerical results, while the dashed line is the analytical approximation $\tl{\psi}_{\rm 
Hall}=4\tl{B}/3$, highlighting the near linear dependence of the Hall potential on magnetic field 
strength.
}
\label{fig_psi_vs_x}
\end{figure}

Figure \ref{fig_phi_vs_x} shows the profiles $\phi(\tl{x})$ as obtained numerically from Eq. 
(\ref{eq_poiseuille_salt}) with varying values of $\tl{B}$ (see legend). The dimensionless length 
$\tl{x}$ is in the range $-\tl{d}\leq\tl{x}\leq \tl{d}$ with $\tl{d}=10$; curves are shown for $\tl{x}>0$ 
only since $\phi(\tl{x})$ is anti-symmetric in $\tl{x}$. All curves start with $\phi(\tl{x}=0)=0$ (the 
potential in the midplane is zero) and decrease with distance. The surface value $\phi(\tl{d})$ is 
negative and decreases with $\tl{B}$.

The nonlinear boundary value problem was solved using a standard shooting method. We 
utilized a Runge-Kutta scheme to integrate the ordinary differential equation from one boundary, 
while an iterative root-finding algorithm adjusted the initial conditions to satisfy the boundary 
condition at the opposite wall. To evaluate the numerical error for a given profile $\phi(\tl{x})$, we 
wrote Eq. (\ref{eq_poiseuille_salt}) in the form $LHS=RHS$, where left- and right-hand side are the 
two sides of the equation. The relative error is defined as $err=|RHS-LHS|/(|RHS|+\epsilon_{\rm 
mach})$, where $\epsilon_{\rm mach}$ is the machine precision. This prevents division by zero in 
regions where the right-hand side vanishes. For each  calculated curve, $|{\rm err}|$ was smaller 
than the given threshold (typically $10^{-6}$). The errors at the boundaries, $|\phi(\tl{x}=0)|$ and 
$|\phi'(\tl{x}=\tl{d})|$, were similarly small.

The dimensionless potential $\tl{\psi}$ is related to $\phi$ via Eq. (\ref{eq_phi_psi}), and is shown in 
Fig. \ref{fig_psi_vs_x} (a) for the same values of $\tl{B}$ and $\tl{d}$. Potential $\tl{\psi}$ increases 
with the distance from the midplane. We stress that the electrostatic potential $\tl{\psi}$ is {\it not} 
screened at the walls. The magnetic field exists everywhere within the gap between the surfaces 
and hence there is an electric field everywhere, not just at the surface. This is clearly seen as an 
almost linear increase in $\tl{\psi}$ near the midplane ($\tl{x}=0$), indicating a constant electric 
field $E=-\psi'$. 

Figure \ref{fig_psi_vs_x} (b) shows the Hall potential $\tl{\psi}_{\rm Hall}$ as a function of 
magnetic field strength. The Hall potential increase monotonically with $\tl{B}$. The dashed line is 
the approximation $\tl{\psi}_{\rm Hall}=4\tl{B}/3$ [see Eq. (\ref{eq_psid_approx})].

To find the dependence of $\lambda$ on $\tl{B}$ for small $\tl{B}$ values, we use the conservation 
integral $\lambda\int_0^{\tl{d}}\cosh(\phi)d\tl{x}=\tl{d}$ with $\cosh(\phi)\approx 1+(1/2)\phi^2$ for 
$|\phi|\ll 1$, and $\phi$ from Eq. (\ref{eq_phi_ana}). To lowest order this yields
\begin{\ea}\label{eq_lam_ana_approx}
\lambda \approx \frac23\left(\frac{\tl{B}}{\tl{d}^2}\right)^2,~~~~
\lambda\approx1-\frac23\left(\frac{\tl{B}}{\tl{d}^2}\right)^2.
\end{\ea}
The larger of the two values is the correct one, tending to unity in the limit $\tl{B}\to 0$. The 
smaller $\lambda$ tends to zero in the same limit and cannot satisfy the material conservation 
integral in Eq. (\ref{eq_poiseuille_salt}).

Although the influence of magnetic fields is often weak, it is instructive to look at the limit of 
exceedingly large fields. $\phi(\tl{d})$ is negative and decreases with $\tl{B}$. In the limit where 
$\tl{B}\gg \tl{d}^2$, $\phi(\tl{d})$ is linear in $\tl{B}$: $\phi(\tl{d})\approx c_1-c_2\tl{B}$, where $c_1$ 
and $c_2$ are positive constants. The ion conservation integral 
$\lambda\int_0^{\tl{d}}\cosh(\phi)d\tl{x}=\tl{d}$ in Eq. (\ref{eq_poiseuille_salt}) is then 
dominated by $\cosh(\phi(\tl{d}))$, leading to 
\begin{\ea}\label{eq_lambda_large_Bt}
\lambda\approx \frac{1}{\alpha \cosh(\phi(\tl{d}))},
\end{\ea}
where $0\leq\alpha\leq 1$ is a dimensionless constant. That is, $\lambda$ is exponentially small in 
$\tl{B}$. 
\begin{figure}
\centering
\includegraphics[width=\columnwidth,clip]{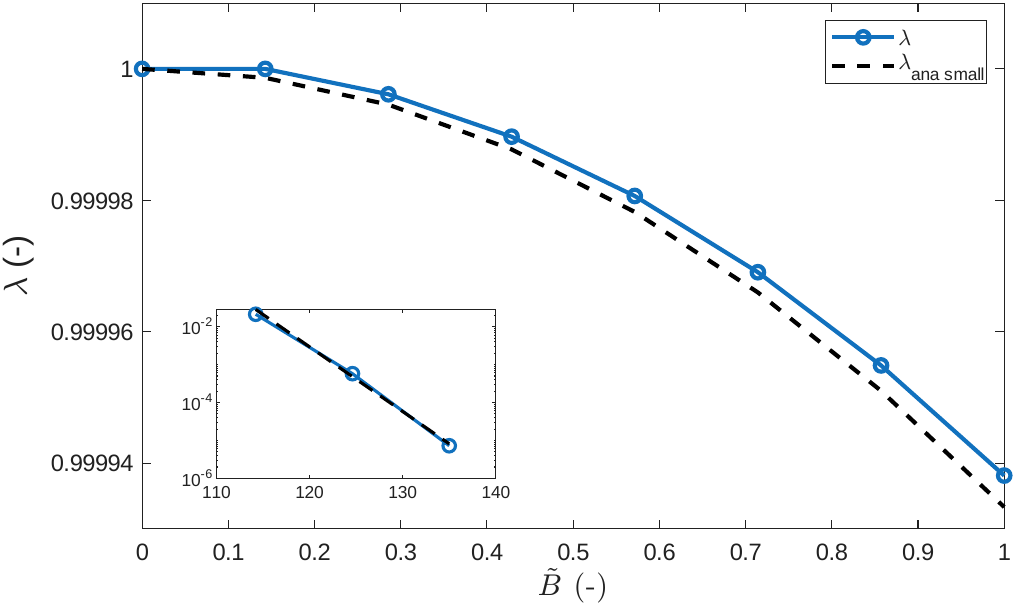}
\caption{The nonlinear eigenvalue $\lambda$ versus scaled magnetic field $\tl{B}$ for 
the same curves shown in Figs \ref{fig_phi_vs_x} and \ref{fig_psi_vs_x}. 
The solid line shows numerical values obtained from the Poisson-Boltzmann equation, while the 
dashed line is the small-$\tl{B}$ approximation from Eq.~(\ref{eq_lam_ana_approx}). 
The inset shows a semi-log plot of $\lambda$ at very large $\tl{B}$, revealing an exponential 
decay. Physically, decreasing $\lambda$ indicates ion depletion in the midplane and enhanced 
accumulation near the walls.  Line with symbols is $\lambda$, dashed line is from Eq. 
(\ref{eq_lambda_large_Bt}).
}
\label{fig_lambda_vs_Bt}
\end{figure}

The eigenvalue $\lambda$ is plotted in Fig. \ref{fig_lambda_vs_Bt} as a function of magnetic field 
strength. When $\tl{B}=0$, there are no forces on ions and they are evenly distributed in the 
electrolytes. Hence their midplane density is $n_m=n_0$ and $\lambda=1$. As $\tl{B}$ is increased, 
ions are pushed toward the walls and depleted from the midplane; $\lambda$ correspondingly 
decreases. For small enough values of $\tl{B}$, $\lambda$ decreases as a parabola, and the 
approximation of Eq. (\ref{eq_lam_ana_approx}) (dashed line) holds well. At large values of $\tl{B}$, 
$\lambda$ levels off, as it cannot be negative. The large $\tl{B}$ limit is shown as a semi-log plot in the 
inset. The straight line is a manifestation of the exponential dependence of $\lambda$ on $\tl{B}$. The 
dashed line is the approximation in Eq. (\ref{eq_lambda_large_Bt}).

\begin{figure}
\centering
\includegraphics[width=\columnwidth,clip]{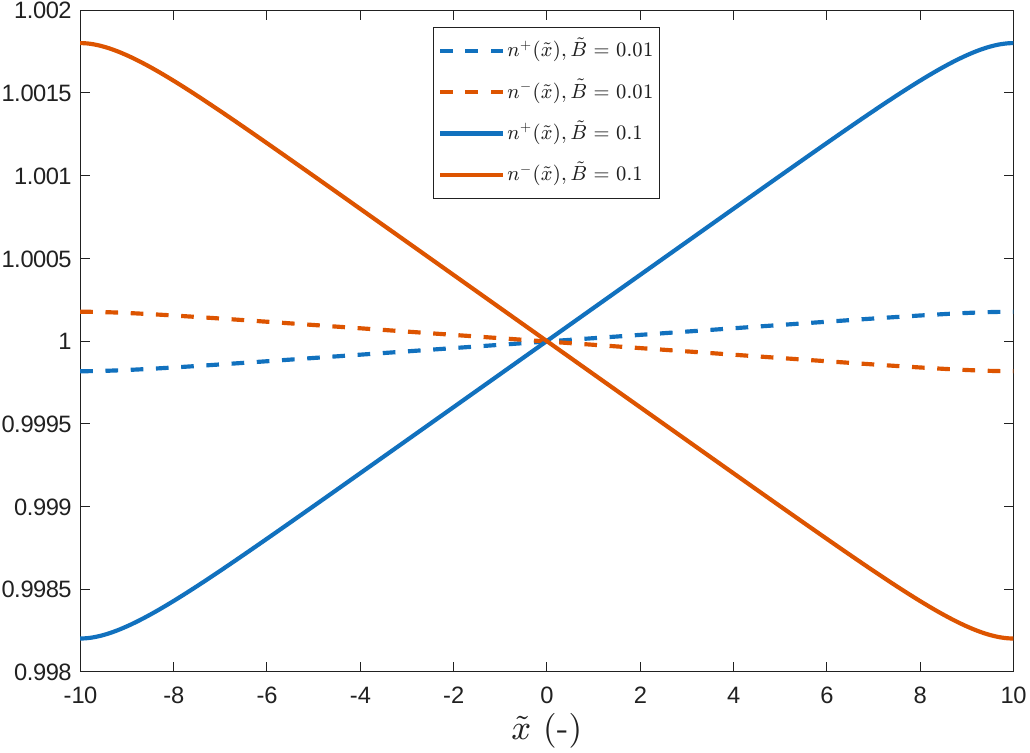}
\caption{Cation and anion number densities $n^\pm(\tl{x})=\lambda\exp(\mp\phi)$ for two 
values of $\tl{B}$. The curves show how magnetic field alters the ion distribution: at higher
$\tl{B}$, ions are pushed toward the walls, leading to long-range separation and deviation from 
classical electric double layer behavior.
}
\label{fig_npm_vs_x}
\end{figure}

The ionic densities are plotted in Fig. \ref{fig_npm_vs_x} for the whole range 
$-\tl{d}\leq\tl{x}\leq \tl{d}$. The vertical axis is $n^+=\lambda\exp(-\phi)$ for cations and 
$n^-=\lambda\exp(\phi)$ for anions. The dashed lines are for weak magnetic field while the solid lines 
are for the large $\tl{B}$ value (see legend). These curves do {\it not} show exponential screening 
close to the surface at $\tl{x}=\pm\tl{d}$. Rather, they vary over macroscopic distance, comparable to 
the slit's width.

\section{Two-dimensional entrance effect and spatial transients}
\label{sec:entrance_effect}

To validate the fully developed unidirectional flow assumption utilized in Secs. \ref{sec_theory} and 
\ref{sec_poiseuille}, and to analytically estimate the entrance length required for the flow to 
become fully developed, we analyze the spatial transients near the entrance of the magnetic 
field region. We assume a unidirectional flow entering from the $y<0$ region where ${\bf B}=0$ 
into the $y> 0$ region where the magnetic field is acting. While the flow at the immediate inlet 
may deviate significantly from the fully developed state, we can extract the dominant spatial decay 
lengths governing the relaxation by performing an asymptotic spatial stability analysis. To do this, 
we linearize the ion densities $n^\pm$, the electric potential $\psi$, and the velocity field ${\bf u}$ 
around the downstream unidirectional analytical solutions of the infinite channel derived in the 
previous sections, denoted by the subscript ``1''. The base states satisfy the one-dimensional 
equilibrium inside the channel ($y>0$). Specifically, the base velocity is the Poiseuille profile ${\bf 
u}_1(x)=v_0[1-(x/d)^2]\hat{y}$. The base ion densities are $n^\pm_1(x) = \lambda n_0 
e^{\mp\phi_{1}(x)}$, and the normalized base potential is $q\psi_1/k_BT \approx 
\tl{B}[x/d-(x/d)^3/3]$. The small two-dimensional spatial perturbations are denoted by $\delta$:
\begin{eqnarray}
    n^\pm(x,y) &=& n^\pm_1(x) + \delta n^\pm(x,y),\nn \\
    \psi(x,y) &=& \psi_1(x) + \delta\psi(x,y),\nn \\
    {\bf u}(x,y) &=& {\bf u}_1(x) + \delta {\bf u}(x,y).
\end{eqnarray}

Substituting these decompositions into the Nernst-Planck expression 
Eq. (\ref{fluxes}), the first-order perturbation flux for the cations, $\delta{\bf h}^+$, is given by:
\begin{eqnarray}
    \delta {\bf h}^+ &=& -D\nabla \delta n^+ - q\mu\delta n^+\nabla\psi_1 - q\mu 
    n^+_1\nabla\delta\psi + \delta n^+{\bf u}_1 + n_1^+\delta{\bf u} \nn \\
    &&+ \mu q\left[\delta n^+({\bf u}_1 \times B_0\hat{z}) + n_1^+(\delta{\bf u} \times 
    B_0\hat{z})\right].
\end{eqnarray}
We further simplify this expression by exploiting the separation of time scales between 
momentum and mass transport. In aqueous electrolytes, the Schmidt number, $Sc = \eta/(\rho D)$, 
is typically very large ($Sc \sim 10^3$). This implies that the 
hydrodynamic profile establishes itself orders of magnitude faster than the concentration profile. 
We therefore assume $\delta {\bf u} \approx 0$ and focus purely on the mass transport relaxation 
governed by $\nabla \cdot \delta{\bf h}^\pm = 0$.

The transverse component of the perturbed cation flux can now be written as:
\begin{equation}
    \delta h_x^+ = -D \frac{\partial \delta n^+}{\partial x} + \mu q \delta n^+ \left( v_1(x) B_0 - 
    \frac{\partial \psi_1}{\partial x} \right) - q \mu n_1^+ \frac{\partial \delta \psi}{\partial x}.
\end{equation}
Recall that the analytical base state $n_1^+(x)$ is in transverse equilibrium, meaning that 
$-D \partial n_1^+/\partial x - q \mu n_1^+ \partial \psi_1/\partial x + \mu q n_1^+ 
v_1(x) B_0=0$. Substituting this equilibrium condition back into the perturbation flux simplifies 
$\delta h_x^+$ to:
\begin{equation}
    \delta h_x^+ = -D \frac{\partial \delta n^+}{\partial x} + D \delta n^+ \frac{\partial (\ln 
    n_1^+)}{\partial x} - q \mu n_1^+ \frac{\partial \delta \psi}{\partial x}.
\end{equation}
This confirms that the mass transport perturbation is driven by the mismatch between local 
gradients, rather than directly by the magnitude of the Lorentz force.

\subsection{Reduction to the Hall-Graetz problem}

We now define the salt concentration perturbation $\delta C = \delta n^+ + \delta n^-$ and the 
charge density perturbation $\delta \rho=\delta n^+ - \delta n^-$. 
To justify the decoupling of the charge density $\delta \rho$, we take the divergence of 
the difference between the linearized cation and anion fluxes, $\nabla \cdot (\delta {\bf h}^+ - 
\delta {\bf h}^-) = 0$. By evaluating the electromigration terms and using the 
Poisson's equation $\nabla^2 \delta \psi=-(q/\veps) \delta \rho$, we obtain a screened transport 
equation for the charge perturbation:
\begin{equation}\label{eq_charge_transport}
    v_1(x) \frac{\partial \delta \rho}{\partial y} = D \nabla^2 \delta \rho - \frac{D}{\lambda_D^2} \delta 
    \rho.
\end{equation}

The charge equation contains a large volumetric sink term $-D \delta\rho/\lambda_D^2$ 
representing dielectric screening. By balancing the convective left-hand side of 
Eq. (\ref{eq_charge_transport}) with this screening sink, we extract the 
characteristic streamwise relaxation length for the charge perturbation: $L_\rho 
\sim v_1 \lambda_D^2/D=d\,\text{P\'e}\, (\lambda_D/d)^2$. Because the 
Debye length is microscopic compared with the channel width, 
this electrostatic adjustment is virtually instantaneous. It is orders of 
magnitude shorter than both the standard hydrodynamic entrance length ($L_{\rm 
hydro} \sim d\, Re$) and the salt concentration Graetz length \citep{bird_stewart_lightfoot_2002}. 
In the macroscopic bulk, this sink overwhelmingly dominates the 
diffusion term $D\nabla^2\delta\rho$, forcing the bulk charge density 
perturbation to vanish almost immediately. At the walls, however, the convective 
left-hand side vanishes ($v_1\to 0$), allowing transverse diffusion to balance 
the screening sink and leading to the creation of an electric double layer. In sum, $\delta \rho$  
does not exhibit a long downstream wake, but is instead confined to these exponential boundary 
layers. In the macroscopic bulk flow, the charge density rapidly converges to its fully developed 
state, making the assumption $\delta \rho\approx 0$ valid almost immediately.

The sum of the linearized fluxes for cations and anions, $\delta {\bf h}_{\rm salt}=\delta {\bf h}^+ + 
\delta {\bf h}^-$, yields:
\begin{equation}
    \delta {\bf h}_{\rm salt}=-D \nabla \delta C + \delta C {\bf u}_1 
    + \mu q \delta \rho ({\bf u}_1 \times B_0\hat{z} - \nabla \psi_1)- 
    \mu q (n_1^+ - n_1^-) \nabla \delta \psi.
\end{equation}
The third term on the right-hand side is vanishingly small in the bulk because the Hall electric 
field establishes itself to oppose the magnetic drift. The fourth term is exponentially small as the 
base ion densities obey $n_1^+ - n_1^- = -2 n_m \sinh(\phi_1)$, and the base potential $\phi_1 \sim 
(\lambda_D/d)^2 \ll 1$. Thus, $n_1^+ \approx n_1^-$. The divergence of the salt flux, $\nabla \cdot 
\delta {\bf h}_{\rm salt}=0$, thus simplifies 
to the standard convection-diffusion balance. Neglecting streamwise diffusion, $\partial_y^2 \ll 
\partial_x^2$, we obtain the governing equation for the salt perturbation in the entrance region:
\begin{equation}
    v_1(x) \frac{\partial \delta C}{\partial y} = D \frac{\partial^2 \delta C}{\partial x^2}.
\end{equation}
This classical Graetz formulation demonstrates that the macroscopic entrance length is dictated 
purely by mass transport scaling, largely independent of the complex electrostatic coupling. 

\subsection{Estimation of the entrance length}

Because the Hall separation is antisymmetric with respect to the midplane, the salt perturbation 
$\delta C$ is an even function of $x$, and its first moment vanishes. To estimate the development 
length, we evaluate the second moment $M_2(y) = \int_{-d}^{d} x^2 \delta C(x,y) \, dx$, 
representing the transverse variance of the concentration. 

Multiplying the Graetz equation by $x^2$ and integrating across the channel, we define a 
variance-weighted effective velocity, $v_{\rm eff} \equiv \int_{-d}^d x^2 v_1 \delta C dx / \int_{-d}^d 
x^2 \delta C dx$. By invoking a self-similar concentration profile, $\delta C(x,y) \approx \delta 
C_{\rm wall}(y) f(x/d)$, the $y$-dependent amplitude cancels out of the ratio, rendering $v_{\rm 
eff}$ a strict constant independent of downstream distance:
\begin{equation}
    v_{\rm eff} \frac{d M_2(y)}{d y} \approx D \int_{-d}^{d} x^2 \frac{\partial^2 \delta C}{\partial x^2} \, 
    dx.
\end{equation}
Integration of the right-hand side by parts twice yields boundary terms and a residual integral. 
The no-flux boundary condition, $\partial \delta C/\partial x=0$ at $x=\pm d$, leads to vanishing of 
the first boundary term. Evaluation of the second boundary term yields $-2d \delta C_{\rm 
wall}(y)$. Crucially, because the Hall effect redistributes ions without creating or destroying mass, 
global salt conservation dictates that the cross-sectional integral of the perturbation must vanish. 
The equation thus reduces to
\begin{equation}
    v_{\rm eff} \frac{d M_2(y)}{d y} \approx -4D d \cdot \delta C_{\rm wall}(y).
\end{equation}
Assuming a self-similar profile where the moment scales with the wall amplitude, $M_2 \approx 
\gamma d^3 \delta C_{\rm wall}$, where $\gamma$ is dimensionless number of order unity, we 
extract a characteristic decay constant. The development length scales as:
\begin{equation}
    L_{\rm dev} \sim \frac{v_{\rm eff} d^2}{D} =d\, \text{P\'e}
\end{equation}
where $\text{P\'e} = v_{\rm eff} d / D$ is the P\'eclet number.

The above analysis reveals a separation of length scales, demonstrating that the 
entrance region operates in two distinct kinematic stages:

\begin{enumerate}
    \item \textbf{The Incompressible stage (electrostatic adjustment, $L \sim d$):} 
    Immediately upstream and downstream of the point where the magnetic window starts acting 
    ($|y| \lesssim 2d$), the concentration gradients are negligible ($\delta n \approx 0$). 
    In this region, the ion cloud behaves as an incompressible fluid. Because the bulk charge density 
    has already relaxed to its fully developed profile ($\delta \rho \approx 0$), Poisson's equation 
    reduces to Laplace's equation $\nabla^2 \delta \psi \approx 0$ for the perturbation potential.
    Solutions to Laplace's equation in a strip dictate that the Hall 
    potential $\psi$ generates fringing electric fields that decay over the geometric scale of the 
    channel width, $y \sim d$. Thus, the fields adjust to satisfy current conservation almost 
    instantaneously, on the scale of the channel width, pre-accelerating ions transversely before 
    they fully enter the magnetic region.
    
    \item {\bf The compressible stage (concentration development,} $L_{\rm dev} \sim \text{P\'e} 
    \,d${\bf ):} Over a much longer length scale, the transverse drift creates accumulation at the walls. 
    In this second stage, the ion cloud becomes ``compressible,'' and the flux balance shifts to include 
    the opposing force of diffusion. This concentration wake extends downstream over a 
    macroscopic scale dictated by the P\'eclet number. 
\end{enumerate}

Therefore, while the minor amplitude modulations of the salt concentration require a long Graetz 
length to fully develop, the electric field and charge density adjust to the magnetic field almost 
instantaneously.

\section{“Hall effect” in Poiseuille flow with only counterions}\label{hall_ci}

It is interesting to look at the ``Hall effect'' when there is no salt in the liquid. Rather, 
the walls of the slit become charged and release the counterions into the liquid. 
Many solid surfaces become charged when immersed in a polar solvent (like water) due to 
surface ionization, ion adsorption, or dissociation of surface groups. For example, 
in Silica (SiO$_2$), surface silanol groups (Si-OH) can deprotonate in water, releasing H$^+$ and 
leaving negatively charged Si-O- groups. Alumina (Al$_2$O$_3$) can both protonate or deprotonate 
depending on pH, releasing either OH$^-$ or H$^+$. In polystyrene sulfonate beads, sulfonic acid 
groups dissociate in water, releasing H$^+$. In carboxylated surfaces, carboxylic acid groups (-COOH) 
deprotonate to -COO-, releasing H. Even clay and minerals, such as montmorillonite and kaolinite, 
have surface hydroxyls and exchangeable cations (Na$^+$, K$^+$, Ca$^{2+}$) that can be released 
into solution.

The walls of the slit are negatively charged with $-q\sigma$ 
Coulomb per unit area. The geometry of the problem is unchanged, only the boundary 
conditions for the electrostatic potential now read
\begin{\ea}
\psi'(x=\pm d)=\mp\frac{q\sigma}{\veps}.
\end{\ea}
In systems where charge regulation applies, this boundary condition will differ; in 
particular, close to the entry to the channel, the boundary condition will depend on the 
coordinate $y$ along the stream.

We use the same formalism and definitions Eqs. (\ref{eq_dimensionless}) and (\ref{eq_phi_psi}) to 
arrive at
\begin{\ea}\label{eq_poiseuille_ci}
\phi''&=&-\lambda e^{-\phi}+\frac{2\tl{B}\tl{x}}{\tl{d}^3},\\
\lambda\int_{-\tl{d}}^{\tl{d}}e^{-\phi}d\tl{x}&=&2\tl{\sigma} 
~,~~~~~~~\phi'(\tl{x}=\pm\tl{d})=\mp \tl{\sigma}.\nn
\end{\ea} 
Equation (\ref{eq_poiseuille_ci}) is nonlinear and inhomogeneous as Eq. 
(\ref{eq_poiseuille_salt}), but here $\tl{\sigma}=l_B^2\sigma$ is the scaled surface density and lengths 
are scaled by the Bjerrum length, defined as
\begin{\ea}
l_B&=&\frac{q^2}{\veps k_BT}.\nn
\end{\ea} 
In water at room temperature, $\veps\approx 80\veps_0$ where $\veps_0$ is the vacuum permittivity, 
$l_B\approx 9$ nm if $q=e$. When the surfaces are charged with single charge in every 
$10$ nm$\times 10$ nm square, the parameter $\tl{\sigma}$ is approximately unity. The nonlinear 
eigenvalue is defined as $\lambda\equiv l_B^3n_L$, where $n_L$ is the counterion density at the 
``left'' plane ($\tl{x}=-\tl{d}$). 

The appearance of only one exponent in the Poisson-Boltzmann equation reflects the existence of 
only cations in the liquid. The absence of mobile anions means the potential 
does not have a defined symmetry with respect to $\tl{x}$. The boundary conditions for the 
Poisson-Boltzmann problem in Eq. (\ref{eq_poiseuille_ci}) are of Neumann type and hence addition of 
a constant to $\phi$ would just renormalize $\lambda$. We may therefore fix the gauge by replacing 
the condition $\phi'(\tl{x}=-\tl{d})=\tl{\sigma}$ by $\phi(\tl{x}=-\tl{d})\equiv 0$, and look for a unique 
value of $\lambda$ without changing the counterion distribution or the physics. 

It is instructive to first look at the solution when $\tl{B}=0$. Eq. 
(\ref{eq_poiseuille_ci}) can then be integrated to give $(1/2)\phi'^2=\lambda e^{-\phi}+c_1$ with a 
constant $c_1$. Potential $\phi(\tl{x})$ is symmetric when $\tl{B}=0$, and using the boundary 
condition, we find
\begin{\ea}
e^{-\phi_m}=1-\frac12\frac{\tl{\sigma}^2}{\lambda},
\end{\ea} 
where $\phi_m\equiv \phi(\tl{x}=0)$ is the midplane potential. Since $e^{-\phi_m}$ is positive, it 
follows that
\begin{\ea}
\lambda>\frac12\tl{\sigma}^2.
\end{\ea} 
For large values of $\tl{\sigma}$, $\phi_m$ is larger than unity, and then $\lambda 
\gtrsim (1/2)\tl{\sigma}^2$. 
\begin{figure}
\centering
\includegraphics[width=\columnwidth,clip]{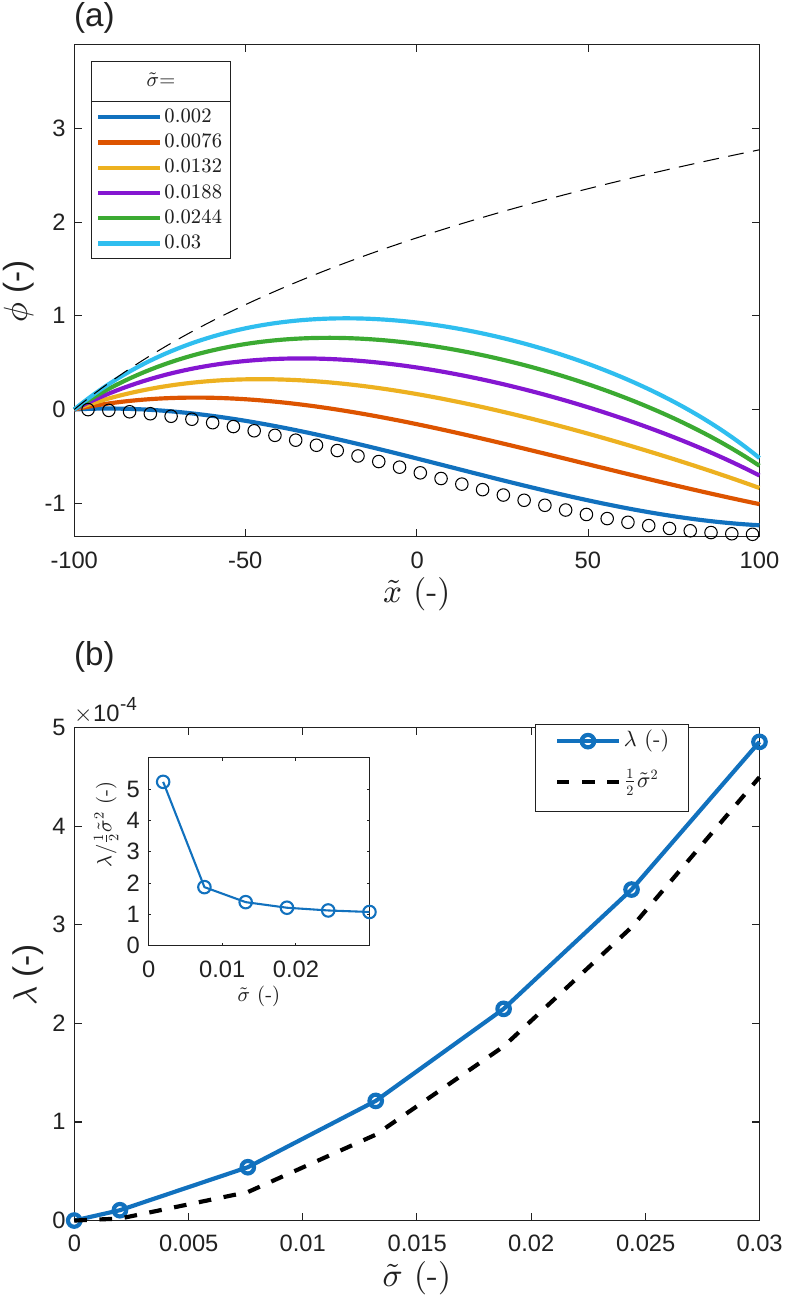}
\caption{(a) Potential $\phi(\tl{x})$ for counterion-only electrolytes flowing in a slit with $\tl{d} = 
100$, for varying surface charge densities $\tl{\sigma}$. Open circles are the analytical 
expression Eq. (\ref{eq_ana_lambda0}) when $\tl{\sigma}=0$ and the dashed line is the analytical 
Eq. 
(\ref{eq_ana_Bt0}) when $\tl{B}=0$. 
(b) Nonlinear eigenvalue $\lambda$ versus $\tl{\sigma}$. The dashed line shows the analytical
approximation $(1/2)\tl{\sigma}^2$, which becomes accurate at high surface charge. 
Inset: ratio $\lambda / (\frac{1}{2} \tl{\sigma}^2)$ approaches unity, confirming the approximation.
}
\label{fig_phi_lam_ci}
\end{figure}

In the absence of field, we recall the analytical solution of Eq. (\ref{eq_poiseuille_ci}) for the 
semi-infinite domain $-\tl{d}\leq\tl{x}\leq 0$ with $\tl{d}\to\infty$. The solution is
\begin{\ea}\label{eq_ana_Bt0}
\phi&=&2\ln\left(\tl{x}+\tl{d}+\tl{b}\right)+\phi_0,\\
\tl{b}&=&\frac{2}{\tl{\sigma}},~~~~~\lambda=\frac12\tl{\sigma}^2,
~~~~~\phi_0=\ln\left(\frac12\lambda\right).\nn
\end{\ea} 
Length $\tl{b}$ is the dimensionless Gouy-Chapman length; in physical units $b=l_B\tl{b}=2\veps 
k_BT/\sigma e^2$, and $\phi_0$ is chosen such that $\phi(\tl{x}=-\tl{d})=0$. The ion density $n^+$ 
varies algebraically: $n^+\sim 2(\tl{x}+\tl{d}+\tl{b})^{-2}$.

A second analytical solution of Eq. (\ref{eq_poiseuille_ci}) exists when $\tl{B}>0$ and the surfaces 
are uncharged: $\tl{\sigma}=0$ and $\lambda=0$. The solution is
\begin{\ea}\label{eq_ana_lambda0}
\phi=\tl{B}\left(\frac13\left(\frac{\tl{x}}{\tl{d}}\right)^3-\frac{\tl{x}}{\tl{d}}\right)-\frac23\tl{B}.
\end{\ea} 
\begin{figure}
\centering
\includegraphics[width=\columnwidth,clip]{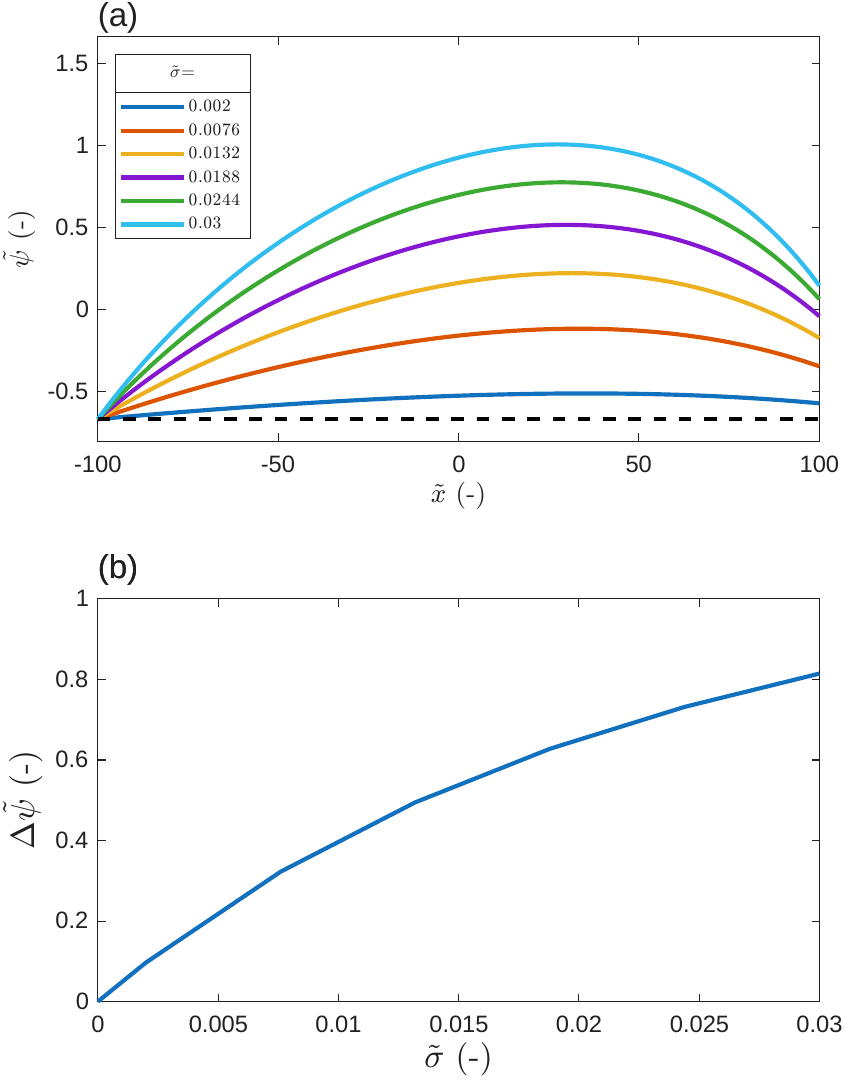}
\caption{(a) Dimensionless electrostatic potential $\tl{\psi}(\tl{x})$ corresponding to the $\phi(\tl{x})$ 
curves in Fig. \ref{fig_phi_lam_ci} (a). Dashed horizontal line is the constant $\tl{\psi}=-2\tl{B}/3$ for the 
no-charge case. As surface charge increases, the potential becomes more concave. 
(b) Potential difference across the slit, $\Delta\tl{\psi}\equiv \tl{\psi}(\tl{d})-\tl{\psi}(-\tl{d})$, 
increases monotonically with surface charge.
}
\label{fig_psi_ci}
\end{figure}

We now turn to the general case where both $\tl{B}$ and $\tl{\sigma}$ are non-zero. Figure 
\ref{fig_phi_lam_ci} (a) shows $\phi(\tl{x})$ for varying values of $\tl{\sigma}$ and fixed $\tl{B}=1$. 
Open circles show $\phi(\tl{x})$ from Eq. (\ref{eq_ana_lambda0}) ($\tl{\sigma}=0$). 
The dashed line is $\phi(\tl{x})$ from Eq. (\ref{eq_ana_Bt0}) with $\tl{B}=0$ and the largest surface 
charge density,  $\tl{\sigma}=0.03$. This value corresponds to one charge on a square of sides $160$ 
nm$\times 160$ nm, which is a low charge density. At these values the effect of magnetic field is 
significant. As $\tl{\sigma}$ increases from zero, the curves deviate from the circles, 
indicating that $\lambda$ increases.

Figure \ref{fig_phi_lam_ci} (b) shows that indeed $\lambda$ increases monotonically with 
$\tl{\sigma}$. Solid line is $\lambda$ obtained numerically from Eq. (\ref{eq_poiseuille_ci}). The 
dashed line is $(1/2)\tl{\sigma}^2$ from Eq. (\ref{eq_ana_Bt0}). This approximation  
for $\lambda$ becomes increasingly accurate for large values of $\tl{\sigma}$ -- the inset 
shows that the ratio $\lambda/\left[(1/2)\tl{\sigma}^2\right]$ tends to unity as 
$\tl{\sigma}$ increases.

Figure \ref{fig_psi_ci} shows the electrostatic potential $\tl{\psi}(\tl{x})$ corresponding to the curves 
in Fig. \ref{fig_phi_lam_ci} (a). When $\tl{\sigma}=0$, $\tl{\psi}=-2\tl{B}/3$ is a constant. With 
increasing surface charge, $\tl{\psi}$ becomes more concave. Fig. \ref{fig_phi_lam_ci}) (b) shows 
the potential difference across the slit: $\Delta\tl{\psi}\equiv \tl{\psi}(\tl{d})-\tl{\psi}(-\tl{d})$. This 
difference varies monotonically with $\tl{\sigma}$. When $\Delta\tl{\psi}=1$, the potential 
difference in physical units is $25$ mV.

\section{Conclusions}\label{sec:conclusions}

We have formulated a theoretical framework for magneto-electrokinetics in dilute electrolytes 
under unidirectional flow, combining the Navier--Stokes, Nernst--Planck, and Poisson equations 
into an analytically tractable system. Unlike classical liquid metals, where charge carriers 
are electrons and conductivity is typically uniform, our model accounts 
for spatially varying ion densities and additional forces such as osmotic pressure.

A key novelty is the sequential decoupling of hydrodynamics and electrostatics: the flow profile 
is governed solely by the Stokes equation, while ion distributions follow a Boltzmann-like law 
incorporating both electrostatic and hydrodynamic potentials. The latter, introduced here, arises 
from the Lorentz force and depends explicitly on the flow profile. The dissolved ions experience an 
algebraically decaying force and the nonlinear Poisson-Boltzmann equation includes this 
location-dependent forcing. As a result, the electric fields are not exponentially screened and can 
be significant even at channel-scale distances from the walls. These insights provide a new 
perspective on electrokinetic phenomena in magnetic fields and establish a foundation for 
analyzing coupled transport processes in electrolytes.

Applying the theory to Poiseuille flow, we derived the Hall potential induced by a transverse 
magnetic field and showed that it scales nearly linearly with flow velocity and magnetic field 
strength, while remaining approximately independent of salt concentration. The analysis also 
extends to counterion-only systems, where surface charge and magnetic forcing jointly shape ion 
distributions. In both cases, the model predicts measurable deviations from classical electrostatic 
screening and identifies a nonlinear eigenvalue $\lambda$ that quantifies midplane ion depletion.

To validate this one-dimensional framework, we analyzed the two-dimensional 
spatial transients near the channel entrance. This analysis revealed a 
hierarchy of length scales, $L_\rho \ll d \ll L_{\rm hydro} \ll L_{\rm dev}$. 
The space charge relaxes almost immediately upon entry, $L_\rho\sim d\, 
\text{P\'e}(\lambda_D/d)^2$. Consequently, the Hall electric field and potential 
adjust over the geometric scale of the channel width, $L \sim d$, governed by 
Laplace's equation. Both of these lengths are much smaller than the standard 
hydrodynamic momentum development length, $L_{\rm hydro}\sim d\, \text{Re}$. In contrast, the 
macroscopic salt concentration wake develops over a much larger Graetz length, $L_{\rm dev}\sim 
d\, \text{P\'e}$, dictated by the P\'eclet number. This confirms that our 
steady-state one-dimensional framework captures the primary 
electrokinetic phenomena for the vast majority of the channel.

Experimental validation of the theoretical predictions is possible. For example, the 
expression for the surface potential $\Delta\psi$ in Eq. (\ref{eq_psid_approx}) can be tested by 
measuring the potential across two sides of the channel with fixed cross-section, while keeping 
$B_0$ constant and varying the pressure drop along the channel, thereby varying the flow rate 
and $v_0$. Alternatively, the flow rate is kept fixed while the magnitude of the magnetic field 
$B_0$ is varied.

The present analysis can be extended to surfaces with charge regulation. For ionization of a 
surface group or adsorption at a single site, \citet{andelman_epl_2016} 
and \citet{levin_jcp_2025} showed how the effective surface charge density 
depends on the reaction's activity and potential. In the present framework, one needs to solve a 
Poisson-Boltzmann equation with boundary conditions that depend in an algebraic way on the 
potential itself. Often, charge-regulation effects are weak so quantities such as the disjoining 
pressure should be examined on distances comparable to the Debye length $\lambda_D$. In 
Sec. \ref{sec_poiseuille} we assumed pure salt and in Sec. \ref{hall_ci} only counterions; 
It would be interesting to study cases where there are both dissolved salt and counterions, as 
even pure water contains some residual mobile charges.

Beyond the Hall effect, the present approach can be generalized to other geometries and flow 
conditions, including Couette and Taylor--Couette systems, and adapted to study electrochemical 
energy devices, microfluidic control, and magnetically-assisted 
desalination \citep{reinke_pof_2018,brockmann_pof_2024}. Future work should 
address nonlinear flow regimes, transient dynamics, and experimental validation, as well as 
explore the interplay between magnetic fields and chemical reactions in confined electrolytes.

\begin{acknowledgments}
This work was supported by the Israel Science Foundation (ISF) grant no. 332/24.
\end{acknowledgments}

\bibliographystyle{apsrev4-2} 
\bibliography{hall_references}

\end{document}